\pdfoutput=1
\documentclass[reprint, aps, prl, superscriptaddress, nobibnotes]{revtex4-1}
\usepackage{amsmath}   
\usepackage{graphicx}
\usepackage{hyperref}
\usepackage[utf8]{inputenc}
\usepackage[english]{babel}
\usepackage[normalem]{ulem}

\begin{document}
\title{Unconventional charge order in a codoped high-Tc superconductor}

\author{D. Pelc}
\email{Correspondence to: mpozek@phy.hr, dpelc@phy.hr}
\affiliation{Department of Physics, Faculty of Science, University of Zagreb, Bijeni\v{c}ka 32, HR-10000, Zagreb, Croatia}

\author{M. Vučković}
\affiliation{Department of Physics, Faculty of Science, University of Zagreb, Bijeni\v{c}ka 32, HR-10000, Zagreb, Croatia}

\author{H.-J. Grafe}
\affiliation{IFW Dresden, Institute for Solid State Research, P.O. Box 270116, D-01171 Dresden, Germany}

\author{S.-H. Baek}
\affiliation{IFW Dresden, Institute for Solid State Research, P.O. Box 270116, D-01171 Dresden, Germany}

\author{M. Požek}
\email{Correspondence to: mpozek@phy.hr, dpelc@phy.hr}
\affiliation{Department of Physics, Faculty of Science, University of Zagreb, Bijeni\v{c}ka 32, HR-10000, Zagreb, Croatia} 

\begin{abstract}

Charge stripe order has recently been established as an important ingredient of the physics of cuprate high-T$_c$ superconductors. However, due to the complex interplay between competing phases and the influence of disorder, it is unclear how it emerges from the parent metallic state. Here we report on the discovery of an unconventional electronic ordered phase between charge-stripe order and (pseudogapped) metal in the cuprate La$_{2-x-y}$Eu$_{y}$Sr$_x$CuO$_4$ with $y = 0.2$ (LESCO). The peculiar properties of the intermediate phase are revealed through three complementary experiments: nuclear quadrupole resonance, nonlinear conductivity, and specific heat. We demonstrate that the order appears through a sharp phase transition, and exists in a dome-shaped region of the phase diagram, similar to charge stripes. A comparison to recent theoretical work shows that the order is a state without broken translational symmetry -- a charge nematic. We thus resolve the complex process of charge stripe development in cuprates, show that the nematic phase is unrelated to high-temperature pseudogap physics, and establish a link with other strongly correlated electronic materials where nematic order is prominent. 

\end{abstract}

\maketitle

Cuprate high-temperature superconductors display a staggering complexity of electronic behaviour which is the subject of intense research and controversy. One of the central questions of cuprate physics is the role of competing electronic ordering tendencies in determining both normal-state and superconducting properties. Recently charge stripe order has emerged as a ubiquitous phenomenon in cuprates\cite{COxray1,stripe_glass2,COxray3,COxray4,YBCONMR1,YBCONMR3}, and charge stripes have been proposed to play a role in almost all of the important features of the cuprate phase diagram: the mysterious pseudogap phase\cite{PG_stripe,nematic_disorder}, Fermi surface reconstruction\cite{LESCO_transport2}, and even superconducting pairing\cite{SC_stripe}. However, firm experimental evidence for these claims is difficult to obtain because charge stripes are sensitive to disorder, which is a prominent feature of all cuprate materials\cite{disorder_rev}. Thus true long-range charge stripe order is never established, and a complex glassy/fluctuating stripe dynamics emerges\cite{stripe_glass1,stripe_glass3,nematic_disorder}. Adding to the intrigue, several theoretical studies predict that charge stripes appear through unconventional precursor phases\cite{electron_nem,nematic_rev,stripe_melt1,stripe_melt2}, such as a charge nematic, which do not break translational symmetry at all. While electron nematic ordering has been discussed extensively in other strongly correlated materials such as pnictide superconductors\cite{pnictide_nem,pnictide_rev} and quantum Hall systems\cite{QH_nematic}, its existence in the cuprates is less clear\cite{YBCO_nematic,BSCCO_nematic,cuprate_nem,nem_rev}. Thus gaining insight into the physics of the emergence of charge stripes from a parent metallic state becomes of great interest.

Here we combine an unusual experimental technique -- nonlinear conductivity -- with two established ones -- nuclear quadrupole resonance (NQR) and specific heat -- to study the appearance of charge stripes in the cuprate La$_{2-x-y}$Eu$_{y}$Sr$_x$CuO$_4$ with $y = 0.2$ (LESCO). Our work provides unprecedented experimental insight into the dynamics and thermodynamics of the process of charge stripe formation in the cuprates. We show that indeed the stripes develop through an unconventional precursor phase consistent with a charge nematic. This phase is found to be insensitive to disorder, appearing at a true thermodynamic phase transition, in agreement with recent theoretical predictions for nematics\cite{nematic_disorder}. In contrast, the transition into the charge stripe phase is smeared out by disorder, resulting in glassiness and short-range stripe correlations. Importantly, we find that the nematic order closely follows the dome-shaped charge stripe appearance in the phase diagram, implying that the high-temperature pseudogap physics is unrelated to nematic/charge stripe occurrence.

We use LESCO as a model system because of its fortunate arrangement of structural and charge-related transitions. It is well known that lanthanum cuprates in their low-temperature tetragonal (LTT) phases have prominent charge and spin stripe order; the archetypal example is La$_{2-x}$Ba$_x$CuO$_4$ (LBCO), where superconductivity is strongly suppressed by stripe order around $x = 1/8$, and separate charge and spin stripe ordering temperatures are observed in neutron scattering, transport and magnetic susceptibility \cite{LBCO_neut1,LBCO_neut2,LBCO_transport}. However, in LBCO the static stripes only exist in the LTT phase and abruptly disappear upon heating above the tetragonal-to-orthorhombic (LTT/LTO) transition, precluding any investigation of their intrinsic melting mechanism in the tetragonal setting. We therefore use the europium co-doped LESCO as a representative system for studying charge stripe physics, due to static charge stripes disappearing spontaneously far below the LTT/LTO transition temperature \cite{LESCO_xray1, LESCO_xray2}. This enables us to reveal the unconventional,  well-defined intermediate phase between stripe order and metal, without interference from structural effects. 

\begin{figure}
\centering
\includegraphics[width=81mm]{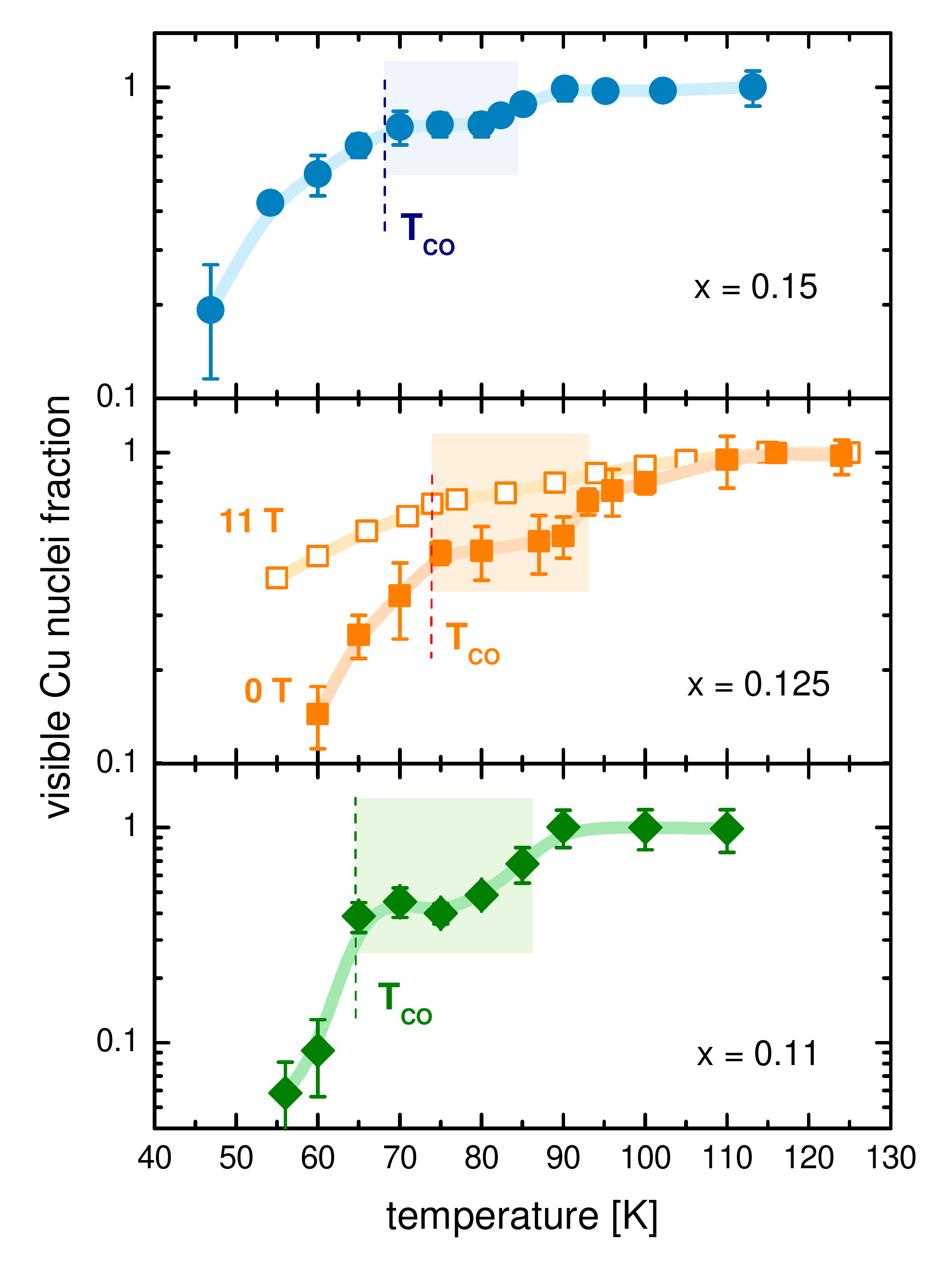}
\caption{\label{LESCO_wipeout} NQR measurements of $^{63}$Cu signal intensity in three LESCO samples, compensated for spin-spin relaxation and Boltzmann temperature dependence, and normalized to high-temperature values. Lines are guides to the eye. The intermediate phase above the charge-ordering temperature is clearly seen as a plateau in the wipeout fraction; note the logarithmic scale. Empty squares for $x = 0.125$ are an additional Cu NMR measurement in an external field of 11~T, also displaying a wipeout plateau. The structural LTO-LTT transition is at $\sim 130$~K for all samples, and the charge stripe ordering temperatures are from a resonant x-ray study\cite{LESCO_xray2}.
}
\end{figure}

\textbf{Nuclear magnetic resonance.} NQR experiments on copper, specifically measurements of NQR signal intensity in dependence on temperature, provide the first indication that an additional phase exists between charge order and the metallic state in LESCO. A decrease of Cu NQR signal -- termed the 'wipeout' effect -- has been previously observed in charge-ordered phases of several cuprates, including LBCO and LESCO\cite{powder_wipeout,wipeout_mag1,wipeout_mag2,LESCO_NQR1,LESCO_NQR2,stripe_glass1}. In our experiments, strong wipeout also begins at the charge ordering temperatures $T_{CO}$ -- which agree with values obtained by resonant X-ray scattering \cite{LESCO_xray2} -- but in addition a wipeout plateau is observed, extending 10 - 20 K above $T_{CO}$ and ending at a temperature $T_{EN}$, significantly below the structural transition (Figure \ref{LESCO_wipeout}). For doping $x = 0.125$ we have performed an additional Cu NMR measurement in a field of 11~T, perpendicular to the crystalline $c$-axis to obtain pure exponential spin-spin relaxation\cite{exprelax}. Due to a measuring frequency $\sim 4$ times higher than in NQR, the NMR wipeout results have significantly better signal-to-noise ratios, and the measurement confirms the existence of an intermediate phase (Fig. \ref{LESCO_wipeout}). Interestingly, the wipeout fraction in 11~T is systematically lower than in pure NQR by a factor of $\sim 1.6$, implying that the strong in-plane field modifies the fluctuations responsible for wipeout. A similar influence is observed\cite{LBCOLa} in La NMR of LBCO-1/8. The detailed investigation of magnetic field effects is left for future discussion.

Although Cu wipeout has been studied before in LESCO and related compounds, to our knowledge the plateau we see here has not been observed before. There are several possible reasons for this: the use of powders instead of single crystals, insufficient resolution, and the dependence of wipeout on magnetic field which has not been systematically investigated. Here both temperature and wipeout resolution is significantly higher than in previous studies, we only work with single crystals to eliminate powder sample/grain boundary issues\cite{powder_wipeout,stripe_glass1,LESCO_NQR1,LESCO_NQR2}, and we have carefully avoided or compensated for spurious effects which could modify the signal intensity. Structural changes cannot influence the results (the LTO/LTT transition is close to $130$~K in the entire investigated doping range\cite{LESCO_xray2,LESCO_mu}), spin-spin relaxation is compensated-for using exponential fits, the NQR lineshape remains essentially the same throughout our temperature range\cite{stripe_glass1,wipeout_mag1}, and skin depth changes at MHz frequencies are of the order of 1\% in our range of temperatures, too small to account for the signal change (see Supplementary Information for more details on these effects and procedures). However, we stress that the microscopic mechanism causing wipeout is a matter of debate, and without a deeper understanding of its origin, wipeout remains a blunt instrument for studying electronic order. Here its main purpose is to show that two distinct characteristic temperatures are always present, allowing us to construct a phase diagram.

\begin{figure}[b]
\centering
\includegraphics[width=85mm]{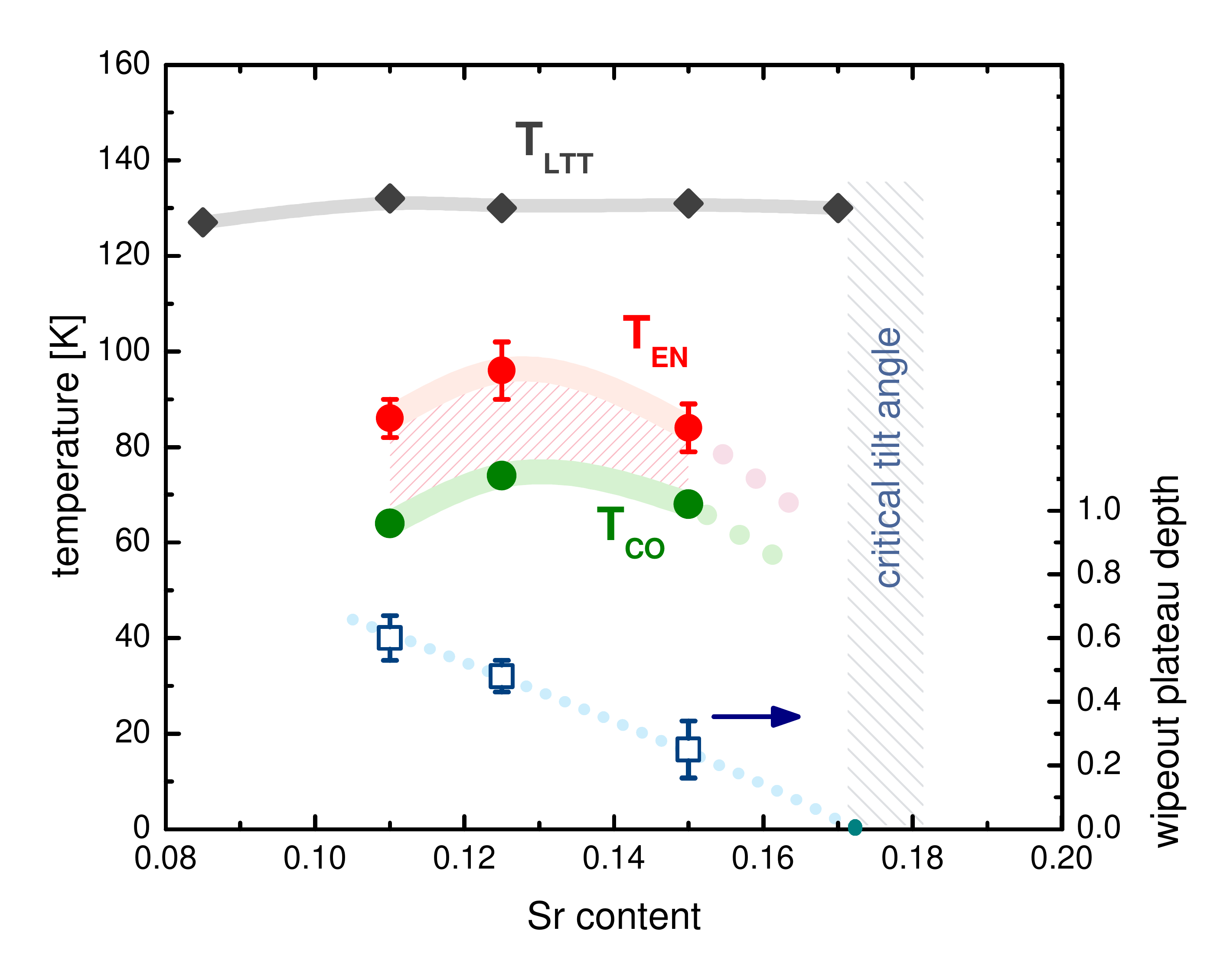}
\caption{\label{LESCO_phase} Phase diagram of LESCO, showing how the intermediate phase, as obtained from NQR wipeout, approximately follows $T_{CO}$ for the samples investigated. The structural transition temperatures $T_{LTT}$ are from Ref. \cite{LESCO_mu}, and significantly higher than either $T_{CO}$ or $T_{EN}$. The height of the wipeout plateau (empty squares) has a different trend with increasing Sr content, indicating that the intermediate phase disappears at a critical doping slightly above $x = 0.17$, which approximately coincides with the critical tilt angle of Cu-O octahedra beyond which the tetragonal phase is unstable.
}
\end{figure}
A close relationship between stripe order and the intermediate phase can be inferred from the behaviour of transition temperatures at different doping concentrations. The intermediate phase transition temperature $T_{EN}$ approximately follows the dependence of $T_{CO}$ on Sr concentration (Figure \ref{LESCO_phase}). Another appealing feature of the wipeout plateau is the dependence of its height (compared to high-temperature values) on doping (Fig. \ref{LESCO_phase}, empty squares), showing a trend different from the transition temperatures. It appears that the intermediate phase either merges with the stripe phase or vanishes altogether at $x \approx 0.17$. The LTT structure becomes unstable at similar dopings \cite{LESCO_mu,LESCO_xray2}, although previous measurements do give indications of stripe order persisting up to $x \sim 0.2$ \cite{powder_wipeout,LESCO_NQR1}. 
 
As a local probe, NQR cannot tell us more about the long-range properties of the intermediate phase, or if it is a distinct state of electronic matter at all.  Since there is no sign of the intermediate phase in X-ray scattering -- which needs broken translational symmetry and coherence lengths of at least a few nanometres -- we conclude that the order is either short-ranged or possesses higher symmetry than charge stripes. This leads to two possible scenarios: phase separation, wherein nanoscale stripe-ordered 'bubbles' are mixed with regions of pseudogapped metal; or a lower symmetry phase ('stripe liquid'), with restored long-range translational symmetry, but rotational symmetry still broken. To decide between the two possibilities we perform two additional experiments on the sample with $x = 1/8$ -- nonlinear conductivity and heat capacity measurements.

\begin{figure}[b]
\centering
\includegraphics[width=86mm]{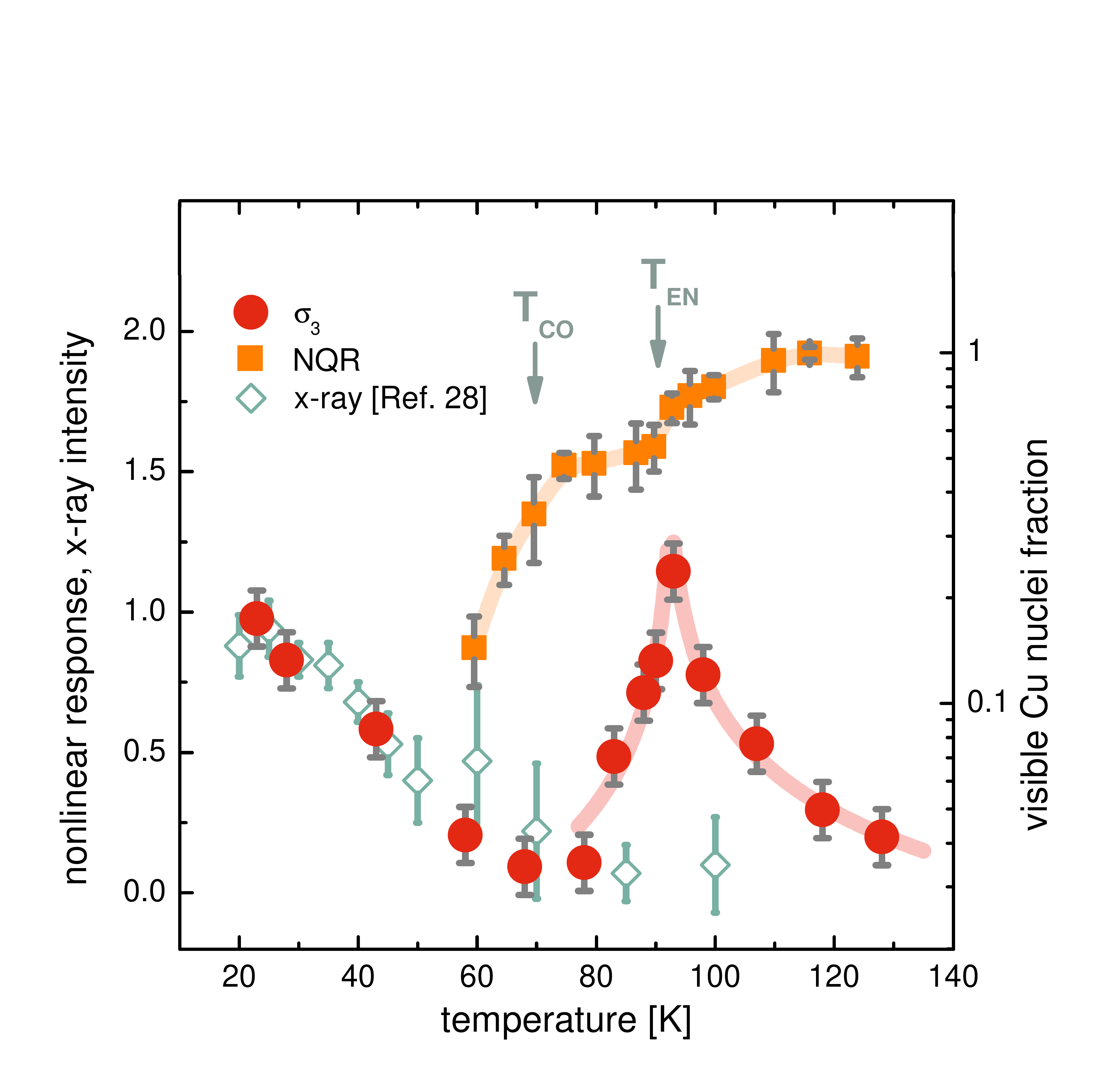}
\caption{\label{LESCO_harm} Third harmonic response in LESCO-1/8 in dependence on temperature (circles), normalized to the low-temperature value. A sharp peak is observed at the intermediate phase transition temperature $T_{EN}$, in good agreement with NQR wipeout measurements (squares). The linear conductivity is featureless. Solid lines are a guide to the eye. The nonlinear signal closely follows the charge stripe peak intensity (diamonds) obtained in a resonant x-ray study\cite{LESCO_xray1} below $T_{CO}$, but not at higher temperatures. This implies that the unconventional phase between $T_{CO}$ and $T_{EN}$ is qualitatively different from charge stripes and does not break translational symmetry.}
\end{figure}
\textbf{Nonlinear conductivity.} Nonlinear conductivity is generally understood as the leading order correction to conventional Ohm's law:
\begin{equation}
\label{sigma3}
j = \sigma E + \sigma_3 E^3 + ...
\end{equation}
where $j$ is the current density, $\sigma$ the linear conductivity, and $\sigma_3$ the leading nonlinear correction -- if time-reversal or inversion symmetry is not broken, only odd powers of $E$ can enter relation (\ref{sigma3}). Due to high linear conductivity of doped cuprates, measuring $\sigma_3$ without significant unwanted heating effects is challenging. We have therefore developed a contactless pulse method \cite{nonlin_mi}, wherein currents are induced in the sample at some frequency $\omega$ (typically $\sim 20$~MHz) and the response at $3 \omega$ (which is proportional to $\sigma_3$) is detected. Results of such a measurement are shown in Fig. \ref{LESCO_harm}, with two characteristic features clearly resolved: charge order below $\sim 75$~K, where a nonlinear signal is expected due to either stripe pinning dynamics \cite{CDW1,CDW2} or glassiness \cite{glass_nonlin,hi4_PRB}, and the dramatic peak at $T_{EN}$. We stress that the \emph{linear} conductivity is smooth and almost featureless (see Supplement). Yet small deviations from high-temperature behaviour were previously detected close to $T_{EN}$ even in linear transport measurements on LESCO \cite{LESCO_transport1,LESCO_transport2} -- possibly another consequence of the unconventional order. 

\begin{figure}
\centering
\includegraphics[width=78mm]{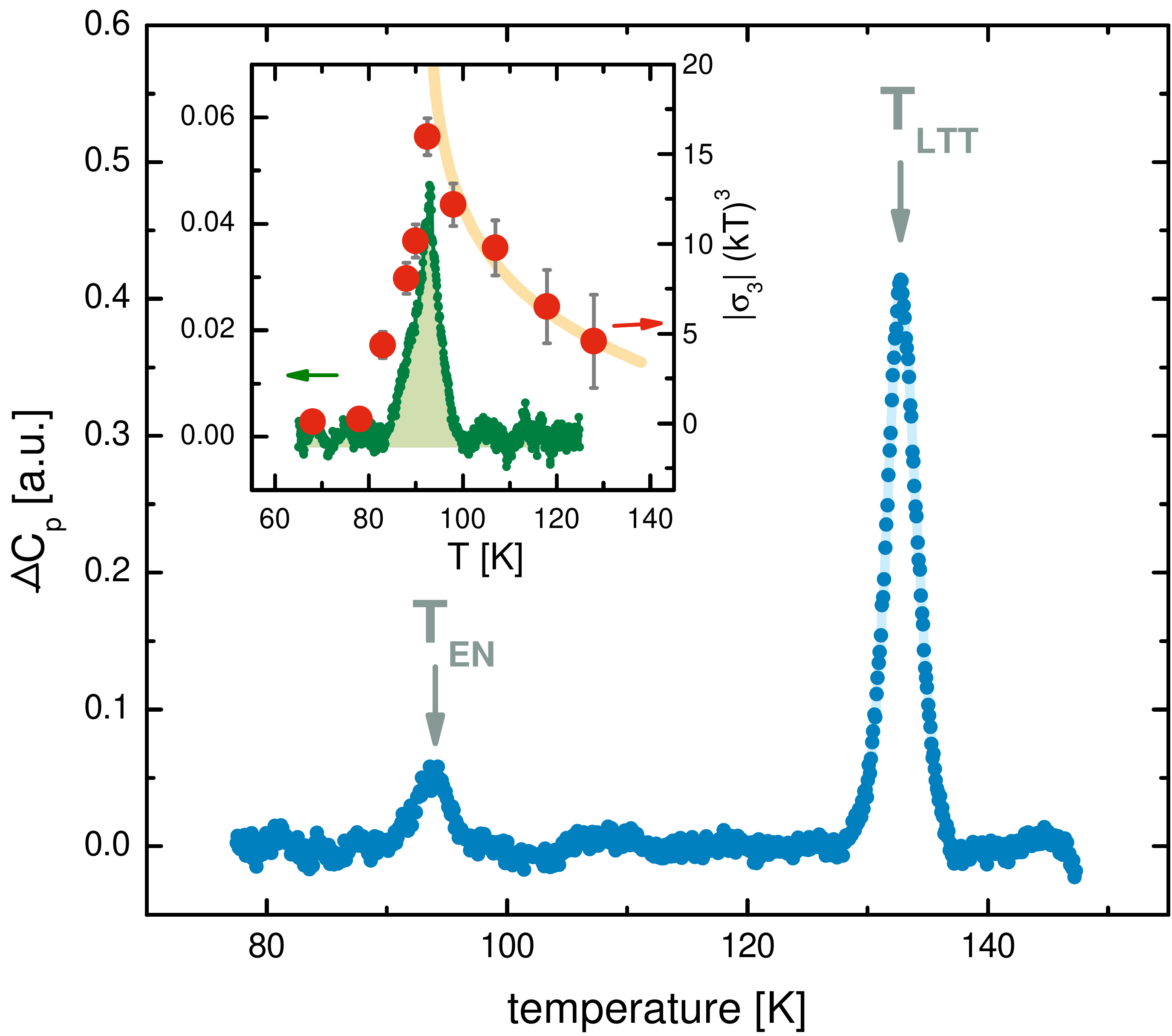}
\caption{\label{LESCO_Cp} Differential heat capacity measurement in LESCO-1/8. Two features can be discerned -- the expected orthorhombic-tetragonal structural transition at $T_{LTT} \sim 130$~K, and a signature of the intermediate phase at $T_{EN}$. The latter is strong evidence that a true thermodynamic phase transition takes place close to $T_{EN}$. Inset shows a comparison between the quadrupolar correlation function $\left| \sigma_3 \right| (kT)^3$ (circles) and $\Delta C_p$ (small circles) for temperatures in the vicinity of $T_{EN}$. The $\Delta C_p$ measurement in the inset is performed $\sim 10$ times slower than in the main graph, enhancing the sharpness of the cusp at $T_{EN}$. The solid line is a fit to a logarithmic divergence $\ln \left| (T-T_{EN})/T_{EN} \right|$, showing that no simple power law (scaling) relation can describe the data. }
\end{figure}

\textbf{Specific heat.} The peak in $\sigma_3$ is highly suggestive of a diverging susceptibility at a phase transition. If a phase transition indeed occurs at $T_{EN}$, it must have a signature in the specific heat, regardless of its microscopic nature. We succeeded in detecting such a signature in a sensitive differential calorimetry experiment. The results are shown in Fig. \ref{LESCO_Cp}, with two visible features -- the structural transition at $T_{LTT} \sim 130$~K, and a distinct peak at $T_{EN}$. A comparison of nonlinear conductivity and $\Delta C_p$ (inset) displays the concordance between transition temperatures. Note that the $\Delta C_p$ in the inset of Fig. \ref{LESCO_Cp} is a different, higher resolution measurement than the one in the main graph, demonstrating both the repeatability and sharpness of the peak at $T_{EN}$. Notably, no sharp features are present at $T_{CO} \sim 75$~K: this is to be expected, since charge stripes are sensitive to disorder and only in a clean material would they occur through a true phase transition. 

\textbf{Discussion.} Our combined experimental approach establishes the existence of a well-defined phase between charge stripe order and the metallic state in LESCO. Furthermore, nonlinear response and specific heat strongly suggest that a distinct type of order exists in the intermediate phase. Namely, the ordering is insensitive to structural disorder present in all doped cuprates, in contrast to stripes: orthorhombic twin domains create pinning centres for stripe order, making it glassy and smearing out the stripe ordering transition \cite{stripe_glass1,stripe_glass2,nematic_disorder}. The temperature $T_{CO}$ is thus not a true phase transition, but rather an onset temperature where translational-symmetry breaking stripe correlations first appear. The intermediate ordering transition, on the contrary, remains thermodynamically well defined. Thus the intermediate phase order clearly behaves in a qualitatively different way from charge stripes, either homogeneous or phase-separated. Although the nonlinear signal close to $T_{EN}$ might also be a dynamical, glassy effect\cite{glass_nonlin}, we expect that this is not the case, for several reasons: the transition is sharp, which would be odd for a glass transition; within experimental sensitivity, we have not observed hysteresis or aging effects characteristic of glassy systems (including spin glasses, which do occur in cuprates at lower temperatures); and the peak-like signature in heat capacity can only mean a true phase transition.

\begin{figure}
\centering
\includegraphics[width=90mm]{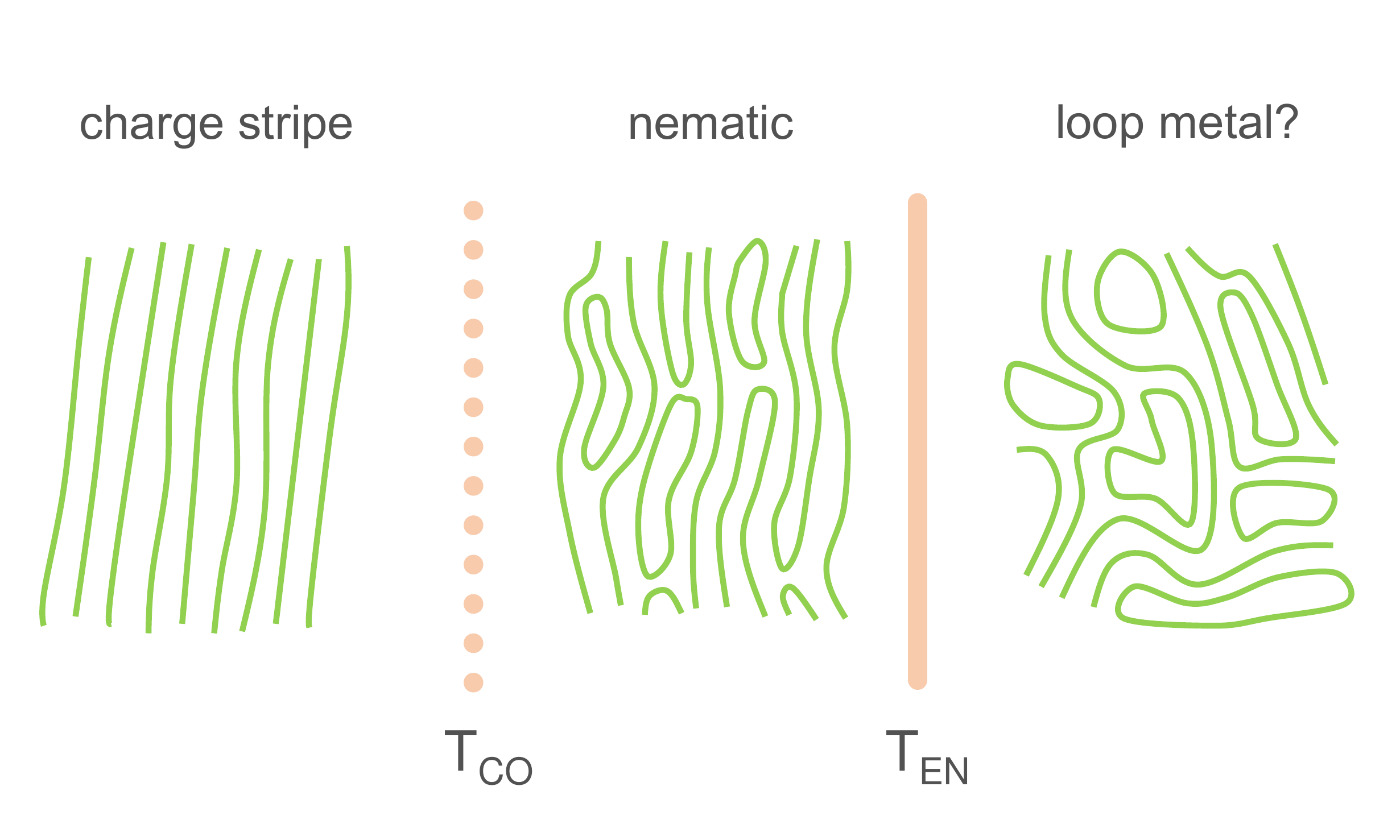}
\caption{\label{sinteza} Electronic liquid crystal physics in LESCO. Three proposed phases are depicted schematically, according to the stripe melting model of Ref. \cite{stripe_melt1,stripe_melt2}: the charge stripe phase below $T_{CO}$, the loop nematic phase between $T_{CO}$ and $T_{EN}$, and the (tentative) loop metal phase with strong nematic fluctuations above $T_{EN}$. Only 'joined' stripe ends exist in both nematic and loop metal phase; other stripe lines are assumed to extend out of the pictures. In a realistic, disordered material, the phase transition at $T_{CO}$ is smeared out and charge stripes become glassy, while the nematic order remains sharply defined.}
\end{figure}

Having established the existence of the intermediate ordering phase, we now show it is consistent with a charge nematic. A recent theory\cite{stripe_melt2} predicts a series of phase transitions between charge stripe order and Fermi liquid (in a clean material), in line with our observations. In that scenario, the intermediate phase would correspond to a nematic stripe loop metal -- a state with broken rotational symmetry, essentially a 'stripe liquid' with proliferated \emph{double} defects (stripe loops), with a preferred direction (see schematic depiction on Fig. \ref{sinteza}). Related nematic order has been discussed in strongly correlated systems \cite{electron_nem,pnictide_nem,pnictide_rev,nematic_rev,YBCO_nematic,BSCCO_nematic,nematic_disorder}. Also, a recent field-theoretical investigation has demonstrated that even weak quenched disorder destroys long-range charge stripe correlations -- precluding a sharp transition in the charge stripe phase -- while the charge nematic transition remains well-defined\cite{nematic_disorder}, in agreement with our data. A charge nematic has been shown to induce spin stripe fluctuations if the system is close to an antiferromagnetic instability \cite{nem_spinfluct}, which is the case in cuprates. Such strong local spin fluctuations provide a natural explanation of the observed wipeout effect\cite{wipeout_mag1,wipeout_mag2}. Normally, the most direct sign of rotational symmetry breaking is the resulting conductivity anisotropy, as e.g. in quantum Hall systems \cite{QH_nematic}, but in the tetragonal phase of LESCO the nematic ordering direction is expected to vary from one CuO$_2$ plane to the next \cite{LBCO_neut2}, so that no macroscopic anisotropy appears in linear conductivity. However, under quite general conditions, an extended fluctuation-dissipation relation connects the quantity $\left| \sigma_3 \right| (kT)^3$ to higher-order, quadrupolar correlations \cite{hi4_PRB}, which have the same symmetry properties as nematic order \cite{degennes,nematic_rev,hi4_PRB}. Thus $\left| \sigma_3 \right| (kT)^3$ is a measure of charge nematic fluctuations. This can be qualitatively understood in the following way: in contrast to the effects of nematic correlations on linear response (which are averaged to zero as described above), nonlinear contributions do not simply average linearly from one CuO$_2$ layer to the next, and a signal will be present as soon as nematic correlations appear. This reasoning is rather general, and similar higher order correlators appear in spin glass/spin nematic physics, where translational symmetry remains unbroken as well. 

Curiously, the temperature dependence of $\left| \sigma_3 \right| (kT)^3$ does not follow any scaling relation (although the small range below $T_{EN}$ precludes any strong conclusions there). Instead, it seems to possess a logarithmic divergence of the form $\left| \sigma_3 \right| (kT)^3 \sim \ln \left| (T-T_{EN})/T_{EN} \right| $ (inset of Fig. \ref{LESCO_Cp}). Although it is difficult to reliably differentiate between a logarithmic dependence and possible corrections to scaling, either way we have an indication that any mean-field description of fluctuations above $T_{EN}$ breaks down. It is tempting to speculate that this is related to the most intriguing prediction of the theory of stripe melting \cite{stripe_melt1,stripe_melt2} -- the existence of another exotic metallic phase between charge nematic and Fermi liquid, a 'stripe loop metal', which should exhibit strong quadrupolar fluctuations \cite{stripe_melt2} and thus large $\sigma_3$.

The results of our NQR, nonlinear response and specific heat investigations taken together with resonant x-ray scattering\cite{LESCO_xray1} enable us to form a complete picture of charge stripe appearance, which we summarize in Fig. \ref{sinteza}. Starting from high temperatures, nematic fluctuations appear and diverge at the well-defined nematic phase transition temperature $T_{EN}$. The nematic phase breaks orientational symmetry and it can be imagined as a stripe liquid with no translational symmetry breaking. This liquid then gradually freezes, with the first appearance of broken translational symmetry at $T_{CO}$ (as seen by x-ray scattering). However, the disorder present in the material strongly influences the stripe freezing, eliminating a sharp charge stripe phase transition and making $T_{CO}$ a crossover temperature. Long range stripe order is never established, as evidenced by relatively small x-ray coherence lengths\cite{LESCO_xray2} and glassy dynamics observed in lanthanum NMR\cite{stripe_glass3,LBCOLa}. 

Our study clearly shows that charge stripes develop through a nematic phase. Since LESCO was carefully chosen as a model material, we expect this scenario to apply to cuprates in general, with some material-specific modifications. Perhaps most notably, the frequently studied YBCO also shows charge stripes in a dome-shaped region\cite{YBCONMR1,COxray1,COxray3}, but the formation of nematic order is more difficult to establish since YBCO already has broken orientational symmetry due to Cu-O chains present in its structure. Furthermore, the recent observation of charge order\cite{COxray4} in the tetragonal HgBa$_2$CuO$_{4+\delta}$ (Hg-1201) might provide an opportunity to further test the phase diagram suggested here. 

Finally, our investigation strongly suggests that the pseudogap is not caused by nematic order. Both nematic and charge stripe ordering tendencies are most prominent around 1/8-doping, while the pseudogap line has a different doping trend and occurs at significantly higher temperatures in the underdoped region. The nematic/charge stripes should thus be viewed as a competing phase with enhanced stability at doping 1/8, with no direct relation to either pseudogap or superconductivity in the cuprates. More broadly, our finding of the unconventional nematic phase in LESCO provides a connection to other materials such as pnictides, and demonstrates the power of using complementary experimental techniques -- local probes like NQR, and unconventional transport measurements, like nonlinear conductivity -- in detecting strange metallic phases. This could be employed in answering fundamental questions in diverse correlated electronic systems, including heavy fermions and quantum magnets.

\textbf{Methods.} For all experiments high quality single crystals of LESCO were used, with typical dimensions of a few mm$^3$. NQR and NMR intensities were obtained by measuring the spin echo signal at the peak of the $^{63}$Cu A line in LESCO \cite{stripe_glass1} and correcting for spin-spin decay (see Supplement for details on the relaxation curves). The entire lineshape was not integrated because it is known that it almost does not depend on temperature (especially in the limited ranges we have in our experiments): the Cu NQR lineshape remains the same all the way from $T_{LTT}$ down to very low temperatures, where it changes drastically due to spin order\cite{stripe_glass1}. This was also confirmed by NMR measurements on our samples (see Supplement). Even if a small broadening of the line occurs, it can not account for the large wipeout -- already in the intermediate phase the signal is a factor of $\approx 2$ smaller than at high temperatures. Spin-spin lattice relaxation was found to be exponential for all samples and relevant temperatures (see Supplement), and was compensated-for using simple exponential fits. The structural transition is close to $130$~K in the entire investigated doping range\cite{LESCO_xray2,LESCO_mu} and thus cannot influence the results. Due to high conductivity, the skin depth in LESCO is below 10~$\mu$m (at the NQR frequency of $\sim 36$MHz) in all investigated samples, making he NQR signal small. We stress that the changes in conductivity at MHz frequencies are of the order of 10\% in our temperature range, too small for the change in skin depth to account for the modified NQR signal.

The nonlinear conductivity setup and measurement procedure are described in detail in Ref. \cite{nonlin_mi}. In experiments on LESCO we have used RF pulses at an excitation frequency of $18.8$~MHz, with pulse lengths of 50~$\mu$s, repetition time 8~ms and maximum pulse power $\sim 100$~W. Measurements were performed with the power varying from 10\% to 100\% full power, and the low power points adjusted to match at all temperatures to avoid baseline drift. Error bars reflect the accuracy of this adjustment. 

Specific heat was measured in a highly sensitive differential thermal analysis (DTA) configuration, wherein the sample was mounted on one of two equal millimeter-sized platinum resistors and the temperature was linearly sweeped. The temperature difference between the resistors with and without sample is then proportional to the sample specific heat. For further details, see Supplementary Information.

\textbf{Acknowledgements.} We thank B. Büchner for providing the LESCO single crystals, N. Bonačić for assistance with the specific heat measurements, and M. Grbić, A. Dulčić and M.-H. Julien for comments. D.P., M.V. and M.P. acknowledge funding by the the Croatian Science Foundation under grant no. IP-11-2013-2729, and S.-H.B. acknowledges support by the DFG Research Grant No. BA 4927/1-1.

\textbf{Author Contributions.} D.P. conceived the study, performed NQR, nonlinear conductivity and specific heat measurements, analysed data and wrote the paper; M.V. designed and constructed the nonlinear conductivity setup and performed measurements; H.-J.G. selected samples and participated in study design and data analysis; S.-H.B. was involved in designing the study and results analysis; M.P. performed NQR measurements and supervised the project. All authors took part in discussing results and editing the manuscript.

\textbf{Author Information.} The authors declare no competing financial interests. Correspondence should be addressed to D.P. (dpelc@phy.hr) or M.P. (mpozek@phy.hr).

\end{document}